\documentclass[aps,amsmath,amsfonts,11pt]{revtex4}
\usepackage{graphicx}
\usepackage{epstopdf}
\usepackage{epsfig,graphicx}
\usepackage[english]{babel}
\usepackage{amsfonts}
\usepackage{amsmath}
\usepackage{latexsym}
\usepackage{graphics,bm}
\usepackage{dcolumn}
\usepackage{bm}
\usepackage{rotating}

\begin{document}

\title{Controllable Normal mode splitting and Switching performance in hybrid optomechanical semiconductor microcavity containing single quantum dot}

\author{ Vijay Bhatt$^{1}$ Sabur A. Barbhuiya$^{2}$ Pradip K. Jha$^{1}$ and Aranya B. Bhattacherjee$^{2}$}

\address{$^{1}$  Department of Physics, DDU College, University of Delhi, New Delhi, 110078, India }
\address{$^{2}$Department of Physics, Birla Institute of Technology and Science, Pilani, Hyderabad Campus, Hyderabad - 500078, India}

\begin{abstract}
	We theoretically explore optical bistability for possible signature of all optical switching and their performance in a hybrid quantum optomechanical system comprising of two semiconductor microcavity coupled optically. One of the cavity is driven by an external optical pump laser while the second cavity which contains a quantum dot is indirectly driven by light transmitted from the first cavity.  The generated bistable behavior due to optomechanical nonlinearity shows a typical optical switching behavior and it can be controlled by changing the laser power, QD cavity coupling, rocking parameter, and the optomechanical coupling. A clear signature of energy exchange between mechanical optical modes is visible from the mechanical displacement spectrum. These results suggest that the present system can be used for an application in sensitive optical switch and optical sensors.
\end{abstract}

\maketitle

\section{INTRODUCTION}
With the aim to construct all optical logic gates by overcoming the obstacles of Moore's law, optical physicists are trying to regulate the transmission of another optical signal using light\citep{saw,rao}. One method to achieve this is optical bistability, in which for a given input power two different output power can be obtained \citep{gibbs}. Thus, the output power is controlled with a shift in the input signal by transition between two stable states. To transfer data over short distances, there is increasing trend of building optical interconnects. In addition these Optical systems could be a fresh computing method \citep{mill,cau} for which bistable optical devices are key component. Optical carrier injection \citep{noz}, thermo-optic effects \citep{alm}, a combination of both \citep{sod}, or optoelectronic feedback \citep{maj} etc. are some methods to achieve bistability. Usually these methods depend on carrier generation which makes it rather slow to use. Second method to get optical bistability is via nonlinear optical effects \citep{kwon}. The nonlinear optical effects in bulk materials are very weak so, a large amount of power is required. This limits the practicality of the device. By enhancing the strength of nonlinear light matter interaction, one can decrease the operating optical power. Taking high quality factor(Q) cavities and small mode volume we can increase the interaction strength.

There has been a major attempt over the previous decade to build solid state optical devices at micro-scale. The primary drive behind this study is the development of chip-to-chip \citep{man1,liu,vla} optical device and the processing \citep{bri,pol,eng1,far1} of quantum information. Modulators play a key role in transmission and routing of data on these networks. In order to minimize operating strength optical components are miniaturized down to scales as such active element is represented by a single quantum emitter. One of the primary obstacles in the development of high-speed electronic devices is currently the big loss in high-frequency metal interconnection. Developing light switches, working at a rate of just a few quanta of energy can be a solution.

Another significant manifestation of strongly coupled system is Normal mode splitting (NMS). It occurs when the energy exchange between two subsystem is faster than it's dissipation to the atmosphere \citep{rossi}. The necessity of this regime is to observe coherent quantum dynamics of the interacting system and to manipulate as well as controlling the quantum system is one of its aim \citep{thomp,wang}. Since the system has inherent Kerr nonlinearity and in addition parametric driving field is introduced, it leads to an interesting concept known as Rocking which essentially transforms a phase invariant self-oscillatory system into the bistable phase\citep{mari,valca}. Where as Kerr nonlinearity weakens NMS, the rocking parameter has been studied to see its effect on controlling it\citep{zhang}.

In this paper we particularly, study the optical bistability for analysing the possible signature of all optical switches and their performance via switch Ratio, Gain and Band width. Specifically, by modeling the dynamics of the bistable switch, we evaluate how the necessary optical power scales with cavity parameters and examine switching speeds. Also, we study about the movable distributed Bragg reflector's (DBR) displacement spectrum which leads to NMS.

\section{THEORY AND MODEL}

The model considered in this article is shown in Fig.1. In the present model, we consider a system comprising of two semiconductor microcavity A and B coupled optically showing two field modes. Cavity A is coupled with movable mirror which is guided using a laser field having frequency $\omega_{p}$. Two level quantum dot is contained in cavity B that is directly driven by the pump field along y-axis with a frequency of $\omega_{L}$ and provide coupling strength $\lambda$. The fabrication of micro-cavities done by a set of distributed Bragg reflectors (DBR). There are some known techniques by which the confinement of light along the longitudinal and transverse direction in the DBR can be achieved \citep{gudat}. DBR mirror comprises of layers of low and high refractive index of quarter-wavelength having AlGaAs as the first and last layer. The refractive index of AlGaAs is greater than that of air and lower than that of GaAs\citep{choy}. Hence, this structure results into an enhancement of the coupling of light.

The optomechanical nonlinearity is introduced into the system through the coupling of the microcavity A to the micro-mechanical resonator. The quantum dot (QD) is considered as two-level system with ground state denoted by $\arrowvert g \rangle$ and excited state by $\arrowvert e \rangle$. 

\begin{figure}[ht]
	\includegraphics [scale=0.5]{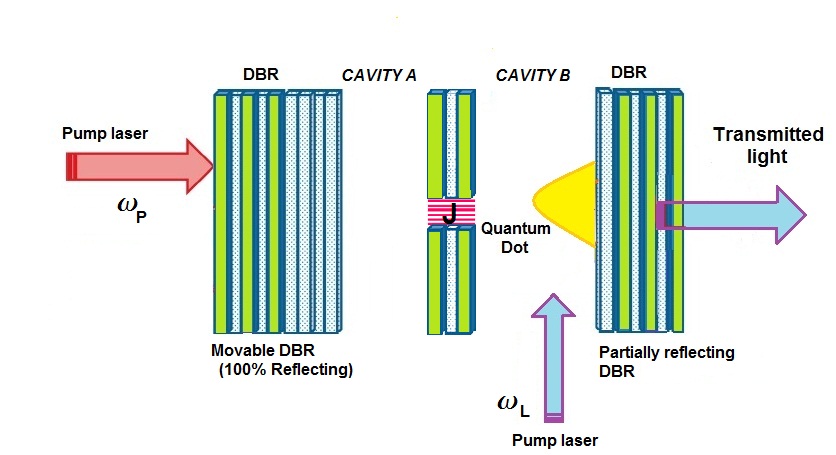}\\
	\caption{ Schematic depiction of the system. DBR mirrors formed two cavity modes coupled with each other. The QD is kept in mode $b$ which interacts with external pump field of frequency $\omega_{L}$. Green strip correspond to AlGaAs and white strip represents GaAs layer.}
\end{figure}

Applying the rotating-wave approximation on the total Hamiltonian for this hybrid optomechanical system along with the dipole approximation, it can be stated as-
 \begin{eqnarray}
H &=& \hbar\Delta_{a}a^{\dagger}a + \hbar\Delta_{b}b^{\dagger}b +\hbar\Delta_{d}\sigma_{z} + \frac{p^{2}}{2m} + \frac{1}{2}m \omega_m^{2}q^{2} + \hbar g(b^{\dagger}\sigma_{ge} + b\sigma_{eg})- \hbar G a^{\dagger}a q \nonumber \\
&+&\hbar J\left(a^{\dagger}b + b^{\dagger}a\right) + i\eta(t) \left(a + a^{\dagger}\right) + \lambda \left(e^{i \delta t+ i\theta}\sigma_{ge} + e^{-i \delta t- i\theta}\sigma_{eg}\right).
\end{eqnarray}

The free energies of cavity A and B optical modes are described by the first and second terms respectively. $a^{\dagger}$ is the creation operator and $a$ is the annihilation operator for cavity A. Similarly $b^{\dagger}$ is the creation operator and $b$ is the annihilation operator for mode B. $\Delta_{a}= \omega_{a}-\omega_{p}$ is detuning for cavity A and $\Delta_{b}=\omega_{b}-\omega_{p}$ is detuning for cavity B with respect to pump laser. Resonance frequencies of both cavity are denoted by $\omega_{a}$ and $\omega_{b}$. The third term shows the energy for the two-level QD. The transition frequency $\omega_{d}$ between two levels of the QD results in its detuning $\Delta_{d} = \omega_{d}-\omega_{L}$. We have taken $\sigma_{ee}$-$\sigma_{gg}$ as $\sigma_{z}$,where $\sigma_{ee}$ and $\sigma_{gg}$ are the atomic populations in the excited and ground levels respectively. Fourth and fifth terms give the mechanical oscillator energy having q and p are position and momentum operator respectively and satisfy [q,p] = i$\hbar$. The interaction in cavity B of the single QD with the optical field is denoted by the sixth term, where g is the coupling strength between QD and photon. The seventh term gives optomechanical interaction, where constant G represents the optomechanical coupling of the mechanical oscillator with field in cavity A. The coupling of two cavities is accounted by the eighth term having coupling strength J between the two cavities \citep{xu,cho,sato,zheng,peng}. Ninth term denotes the driving term with temporally modulated field $\eta(t)=\eta_{0}+P_{amp}Cos(\Omega t)$, where $\eta_{0}$ is constant driving amplitude that takes real value, $P_{amp}$ is the amplitude of the modulation and $\Omega$ is the frequency of modulation. The last term shows the interaction between QD and transverse pump along y-axis, where $\lambda$ is the interaction constant and $\theta$ shows the relative phase between two laser fields.

The system dynamics can be described by the following quantum Langevin equations after using the Hamiltonian resulting from Equation (1):
\begin{equation} 	
\dot{a}(t)=-i\Delta_{a}a-i Jb(t)+\eta(t)+iG a(t)q(t)-\kappa_{a} a(t)+\sqrt{\kappa_{a}}a_{in}(t),
\end{equation}

\begin{equation} 
\dot{b}(t)=-i\Delta_{b}b-i g \sigma_{ge}-iJa-\kappa_{b}b+\sqrt{2\kappa_{b}}b_{in}(t),
\end{equation}

\begin{equation}
\dot{\sigma_{ge}}(t)=(-i\Delta_{d}-\kappa_{d})\sigma_{ge}+i gb(t)\sigma_{z}(t)-i\lambda e^{-i\delta t-i\theta}\sigma_{z}
\end{equation}

\begin{equation}
\dot{q}(t)=\omega_{m}p(t),
\end{equation}

\begin{equation}
\dot{p}(t)=-\omega_{m}q(t)+G a^{\dagger}a-\gamma_{m}p(t)+ \zeta(t),
\end{equation}

 We have assumed $a_{in}(t)$ and $b_{in}(t)$ having zero mean as input vacuum noise operator for cavity A and B respectively. Now behavior of the nonlinear cavity can be analyzed by re-writing the operators in terms of mean classical values.

\begin{equation} 
<\dot{a}(t)>=-i\Delta_{a}<a>-i J<b(t)>+\eta(t)+iG <a(t)><q(t)>-\kappa_{a} <a(t)>+\sqrt{\kappa_{a}}<a_{in}(t)>,
\end{equation}

\begin{equation} 
<\dot{b}(t)>=-i\Delta_{b}<b>-i g <\sigma_{ge}>-iJ<a>-\kappa_{b}<b>+\sqrt{2\kappa_{b}}<b_{in}(t)>,
\end{equation}

\begin{equation}
<\dot{\sigma_{ge}}(t)>=(-i\Delta_{d}-\kappa_{d})<\sigma_{ge}>+i g<b(t)><\sigma_{z}(t)>-i\lambda e^{-i\delta t-i\theta}<\sigma_{z}>,
\end{equation}

\begin{equation}
<\dot{q}(t)>=\omega_{m}<p(t)>,
\end{equation}

\begin{equation}
<\dot{p}(t)>=-\omega_{m}<q(t)>+G <a^{\dagger}><a>-\gamma_{m}<p(t)> + \zeta(t),
\end{equation} 

The interaction between system and external degrees of freedom leads to dissipation. where the decay constants are characterized by $\kappa_{a}$ and $\kappa_{b}$ for the cavity fields A and B respectively and spontaneous emission decay rate of QD is given by $\kappa_{d}$. The $\gamma_{m}$ is the decay constant for mechanical DBR coupled with cavity A.

To analyze further, we consider large and fast modulating input signal,i.e., $P_{amp}>>\eta_{0}$ and $\Omega>> \omega_{q},\Delta_{a},\Delta_{b},\kappa_{a},\kappa_{b},G,\gamma_{m},J$. Further $a$ and $b$ are slowly varying amplitudes and by using few mode expansion\citep{valca}, can be written as $\rightarrow$ $a_{s}+a_{\pm}e^{\pm i\Omega t}$ and b$\rightarrow b_{s}+b_{\pm}e^{\pm i\Omega t}$ respectively. Now putting these expression into equation (7)-(11) and neglecting the higher harmonics and also equating the time derivatives to zero, this gives,

\begin{equation}
b_{s}=\frac{-\left(iJ a + \frac{g\lambda e^{-i\delta t-i\theta} N}{i \Delta_{d}+\kappa_{d}}\right)}{\left(\kappa_{b}+i\Delta_{b}-\frac{g^{2}N}{i\Delta_{d}+\kappa_{d}}\right)},
\end{equation}

\begin{equation}
\sigma_{eg,s}=-i\frac{\left(g b_{s}-\lambda e^{-i\delta t-i\theta}\right)N}{(i\Delta_{d}+\kappa_{d})},
\end{equation}

\begin{equation}
p_{s}= 0,
\end{equation}

\begin{equation}
q_{s}=\chi ( \left|a_{s}\right|^{2}+ C),
\end{equation} 

\begin{equation}
a_{s}=\frac{\eta_{0}\left(A_{1}+iA_{2}\right)+iJg\lambda e^{-i\delta t-i\theta} N}{(i \Delta+ \kappa_{a})(A_{1}+i A_{2})+J^{2}(i\Delta_{d}+\kappa_{d})},
\end{equation}

Hence by eliminating the mode b in the steady state, we get-

\begin{eqnarray}
(A_{1}^{2}
&+ &  A_{2}^2)\omega_{m}^{2}\chi^{4}P_{trans}^{3}-(2(A_{1}^{2}+A_{2}^2)(\Delta_{a}-\omega_{m}\chi^{2}C)\omega_{m}\chi^{2}-2J^{2}\omega_{m}\chi^{2}(\kappa_{a}A_{2}-\Delta_{d}A_{1}))P_{trans}^2\\ \nonumber
&+& ((\kappa_{a}A_{1}+J^{2}\kappa_{d})^2+(\kappa_{a}A_{1}+ J^{2}\Delta_{d})^2+(\Delta_{a}-\omega_{m}\chi^{2}C)^2(A_{1}^{2}+A_{2}^2)\\ \nonumber
&+& 2 J^{2}(\Delta_{a}-\omega_{m}\chi^{2}C)(\Delta_{d}A_{1}-\kappa_{a}A_{2}))P_{trans}\\ \nonumber
&=& |\eta_{0}|^{2}(A_{1}^2+A_{2}^2)+2\eta_{0}A_{2} g J \lambda \cos{\theta} N+J^{2} g^{2}\lambda^{2} N^{2}
\end{eqnarray}

Where $P_{trans}$= $<a_{s}^{\dagger}a_{s}>$ is the power transmitted from fundamental cavity mode A,

Where, $A_{1}=-g^{2}<N>+\kappa_{b}\kappa_{d}-\Delta_{b}\Delta_{d}$,

$A_{2}=\Delta_{b}\kappa_{d}+\kappa_{b}\Delta_{d}$,

$\Delta= \Delta_{a}-\omega_{m}\chi^{2}P_{trans}-\omega_{m}\chi^{2}C$ is the effective detuning in the optomechanical cavity

$C=\frac{P_{amp}^{2}}{2\Omega^{2}}$  is rocking parameter.

$ \chi = G/\omega_{m}$ is the rescaled optomechanical coupling constant.

$P_{amp}$ is the amplitude of the modulating wave and $\Omega$ is the modulating frequency.

Now the dynamics of quantum fluctuations in the system around its stable state will be analyzed. For this, the value of quantum Langevin equations are written as p(t) $\rightarrow p_{s}$+p(t), q(t) $\rightarrow q_{s}$+q(t),  b(t)$\rightarrow b_{s} $+b(t) and a(t) $\rightarrow a_{s}$+a(t) . Here we assume zero fluctuation for operator of QD.

Introducing the quadrature of phase and amplitude for the field, $v_{1}$=i ($b^{\dagger}$-b)/ $\sqrt{2}$, $u_{1}$=(b+$b^{\dagger}$)/ $\sqrt{2}$,$v_{2}$= i($a^{\dagger}$-$a$)/ $\sqrt{2}$, $u_{2}$=($a$+$a^{\dagger}$)/ $\sqrt{2}$, $v_{1in}$= i($b_{in}^{\dagger}$-$b_{in}$)/ $\sqrt{2}$,$u_{1in}$=($b_{in}$+$b_{in}^{\dagger}$)/ $\sqrt{2}$, $v_{2in}$= i($a_{in}^{\dagger}$-$a_{in}$)/ $\sqrt{2}$,$u_{2in}$=($a_{in}$+$a_{in}^{\dagger}$)/ $\sqrt{2}$. Hence the quantum Langevin equations for the quadrature becomes:

\begin{equation} 	
\dot{u_{1}}(t)=\Delta_{b}v_{1}(t)+ J v_{2}(t) -\kappa_{b} u_{1}(t)+\sqrt{\kappa_{b}}u_{1in}(t),
\end{equation}

\begin{equation} 	
\dot{v_{1}}(t)=-\Delta_{b}u_{1}(t)- J u_{2}(t) -\kappa_{b} v_{1}(t)+\sqrt{\kappa_{b}}v_{1in}(t),
\end{equation}

\begin{equation} 	
\dot{u_{2}}(t)=\Delta_{a}v_{2}(t)+ J v_{1}(t) + i a_{-}q(t) -\kappa_{a} u_{2}(t)+\sqrt{\kappa_{a}}u_{2in}(t),
\end{equation}

\begin{equation} 	
\dot{v_{2}}(t)=-\Delta_{a}u_{2}(t)- J u_{1}(t) +  a_{+}q(t) -\kappa_{a} v_{2}(t) + \sqrt{\kappa_{a}}v_{2in}(t),
\end{equation}

\begin{equation} 
\dot{q}(t)= \omega_{m}p(t),
\end{equation}

\begin{equation}
\dot{p}(t)= -\omega_{m}q(t) + a_{+}u_{2}(t) - i a_{-}v_{2}(t) - \gamma_{m}p(t) +\zeta(t),
\end{equation}

Where,

\begin{equation}
a_{\pm}= \frac{\omega_{m}\chi(a_{s} \pm a_{s}^{*})}{\sqrt{2}}
\end{equation}

\section{OPTICAL SWITCH AND IT'S PERFORMANCE}

The vast development of worldwide data and huge demands of communication networks with larger capacity of data transmission and processing information with fast rates requires large bandwidth and low consumption of power\citep{mark,gosh} and for this Optical interconnects is more preferable than electronic interconnects.  The foundation of all-optical switching rests on nonlinear Kerr effect which arises due to refractive index variation of nonlinear material induced by a control light\citep{chai}. The primary interest of optical switching is that it allows the optical data signals to be routed without the need for conversion to electrical signals hence it is not dependent on data protocol or rate\citep{papa}. Therefore all optical switching can taken as a key element, playing a critical and important role in construction on-chip ultrafast all-optical switch networks.

In our model, a bistable device is used as the base to analyze the performance of an optical switch. The input signal consists of a modulated optical signal and a fixed bias power $P_{bias}$.

Using equation(17) we plot the bistability graph between output power and input power (Fig.2). Here we define $P_{trans}$ as output power and $\eta^{2}$ as input power. For N=0 (i.e, equal population in the ground state and excited state) we plot bistability graphics for different values of Rocking parameter (C). 

 \begin{figure}[ht]
	\includegraphics[scale=0.4]{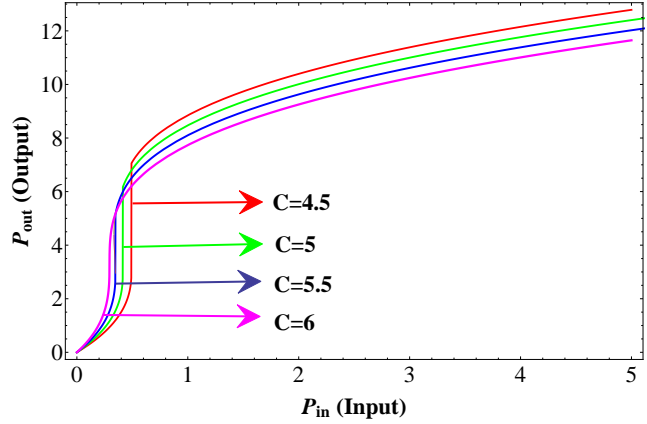}
	\caption{The bistability plot in steady state for different values of C to identify the bias-points, around which the input power can be modulated to observe the output power change; Parameters used for graph are- J=0.5, $\chi=0.3$, $\kappa_{d}=1.8\omega_{m}$, N=0 ,g=1$\omega_{m}$, $\theta=0.238$, $\lambda$=0.02, $\kappa_{a}=0.1\omega_{m}$, $\kappa_{b}=0.1\omega_{m}$, $\Delta_{a}=\omega_{m}$, $\Delta_{b}=\omega_{m}$, $\Delta_{d}$=0.}
\end{figure}

As we increase the value of C the maximum output shift towards lower value of input power. This indicates that the rocking parameter influence the output behavior in a significant manner. Using this, we further illustrate the switch ratio(S.R) and gain of the switch.

During the driving curve of bistability, the switching ratio is defined as the proportion of the maximum to minimum cavity output. Fig.3(a) shows the graph of switching ratio versus input amplitude $P_{amp}$. We find that the switch behaves like a low pass filter. The ratio decreases as the value of $P_{amp}$ increases. Fig.3(b), shows the graph of switching ratio versus output signal as a function of frequency of modulating signal. We observe switching ratio increases with modulation frequency. The larger ratio indicates that switch is controlling the propagating light signal more precisely. This suggests that the input amplitude and modulation frequency influence the switch performance significantly. 

 \begin{figure}[htb]
	\centering
	\begin{tabular}{@{}cccc@{}}
		\includegraphics[scale=0.4]{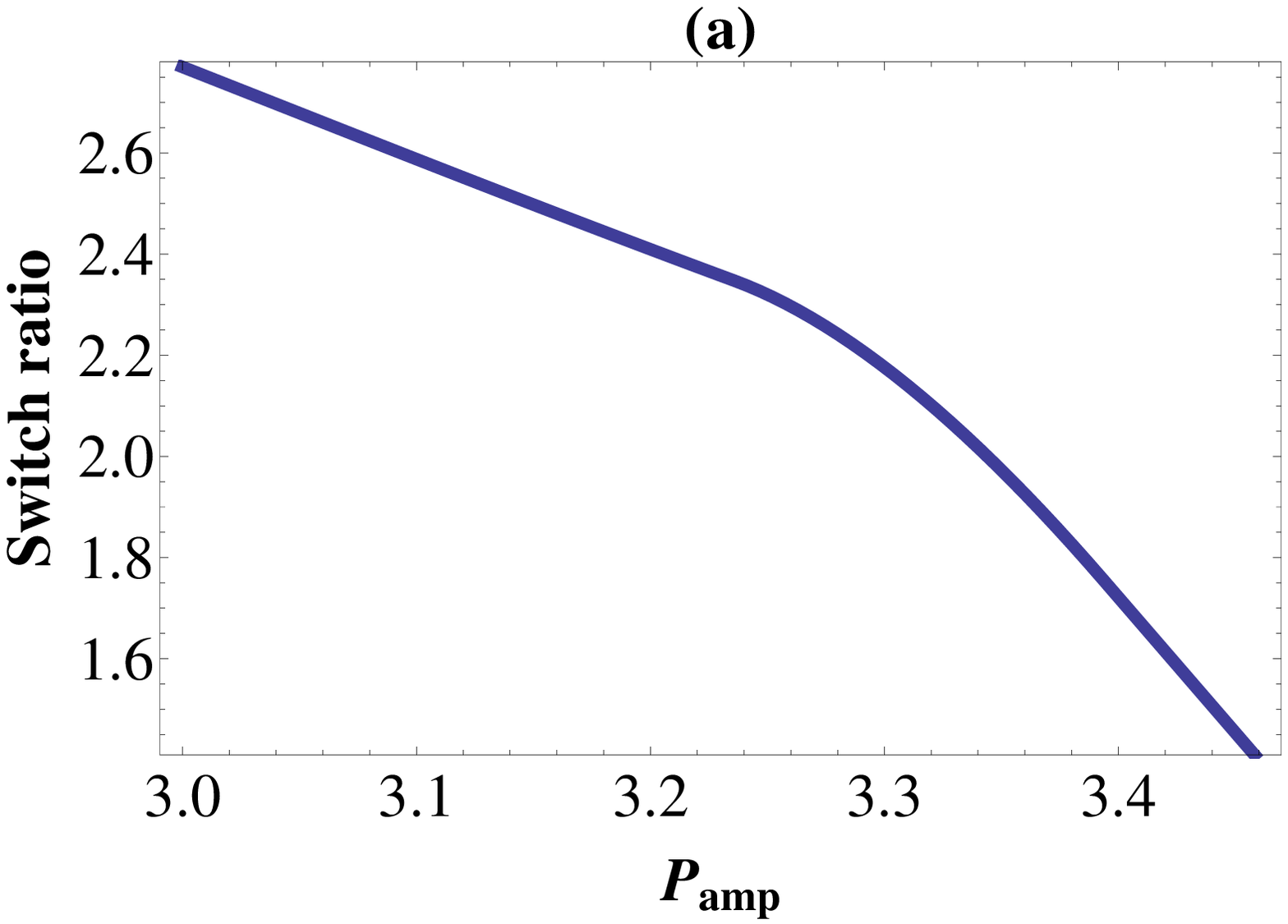}&
		\includegraphics[scale=0.4]{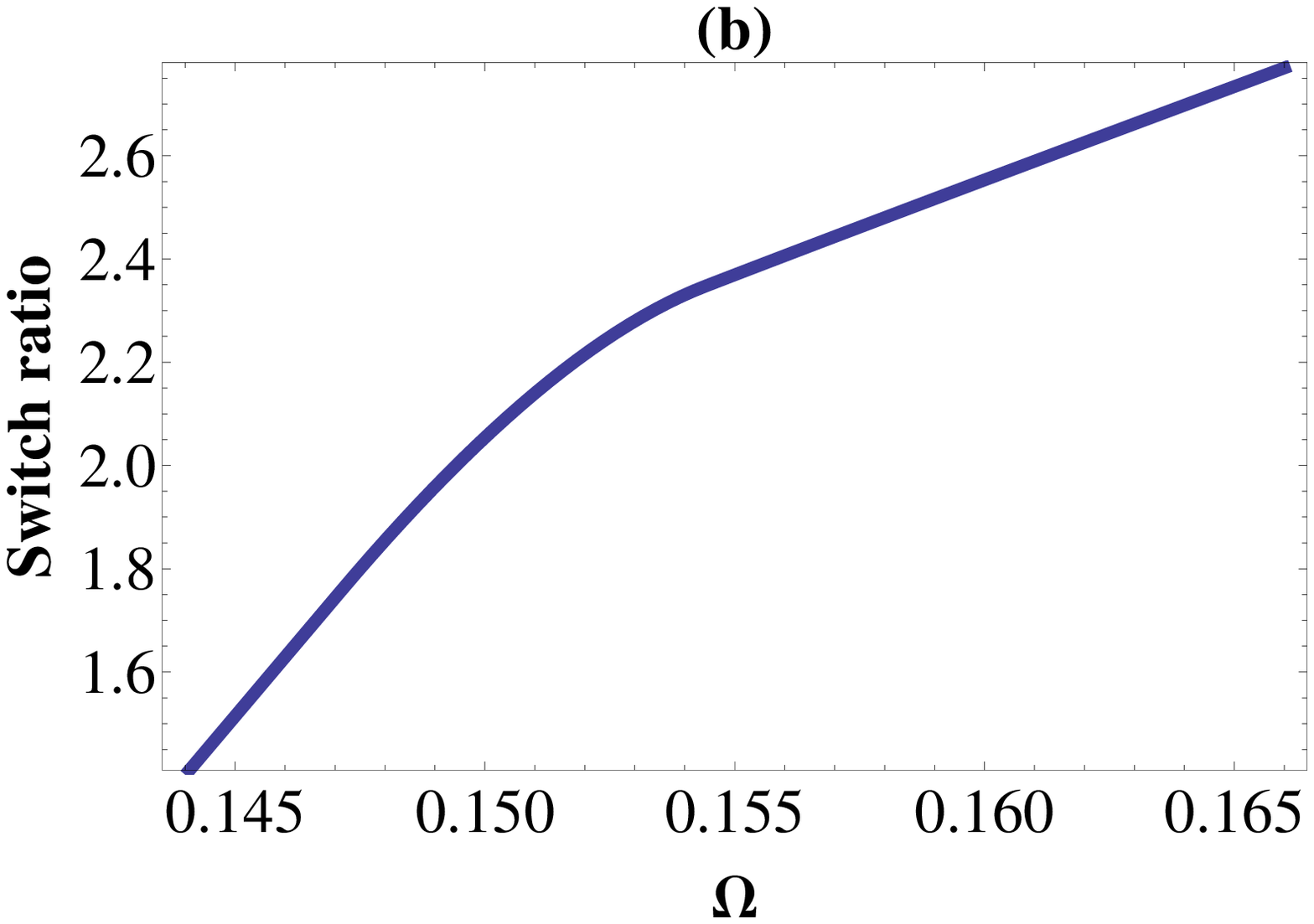} \\
	\end{tabular}
	\caption{The graph of (a)-Switching ratio v/s $P_{amp}$ for fix $\Omega$=1; (b)- Switching ratio v/s Modulation frequency($\Omega$) for fix $P_{amp}=0.5$. Parameters are used for graph are; J=$\omega_{m}$, $\chi=0.3\omega_{m}$, $\gamma_{m}$=1.8$\omega_{m}$, N=0,g=0.5$\omega_{m}$, $\theta=0.238$, $\lambda$=0.02, $\kappa_{a}=0.1\omega_{m}$, $\kappa_{b}=0.1\omega_{m}$, $\Delta_{a}$=$\omega_{m}1$, $\Delta_{b}=1\omega_{m}$, $\Delta_{d}$=0.}
\end{figure}

The Gain of the switch is defined as the ratio of output and input power amplitude. By analyzing gain as a function of  $P_{amp}$ we find that initially the gain value is higher for lower $P_{amp}$ and subsequently the value of gain smoothly decreases as the value of $P_{amp}$ increases (see(Fig. 4(a)). We also notice that from Fig.(4(b)) the gain increases as the value of $\Omega$ increases and than become constant. It may be noted that for a good optical switch a large gain is required. 

\begin{figure}[htb]
	\centering
	\begin{tabular}{@{}cccc@{}}
		\includegraphics[scale=0.4]{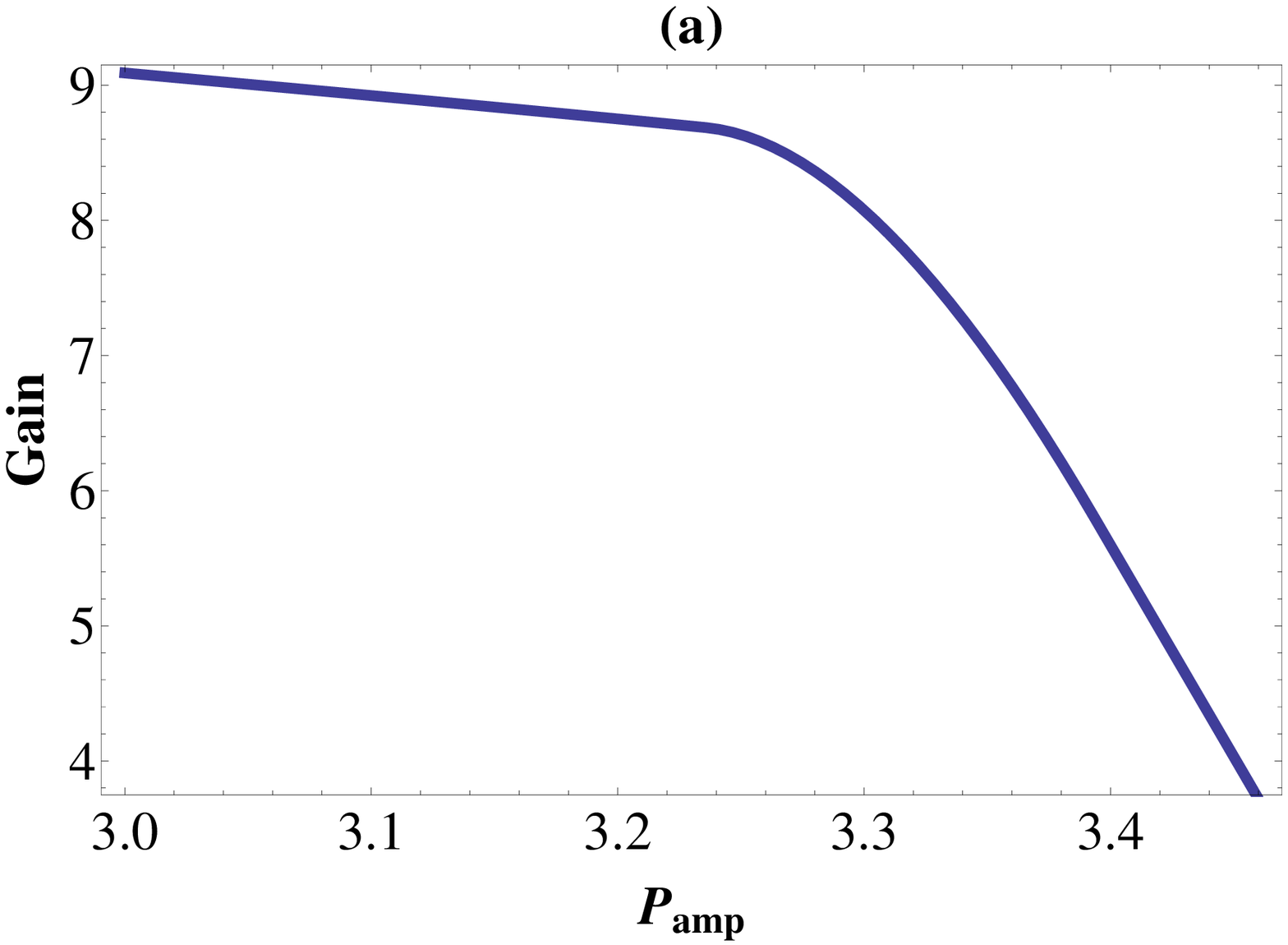}&
		\includegraphics[scale=0.4]{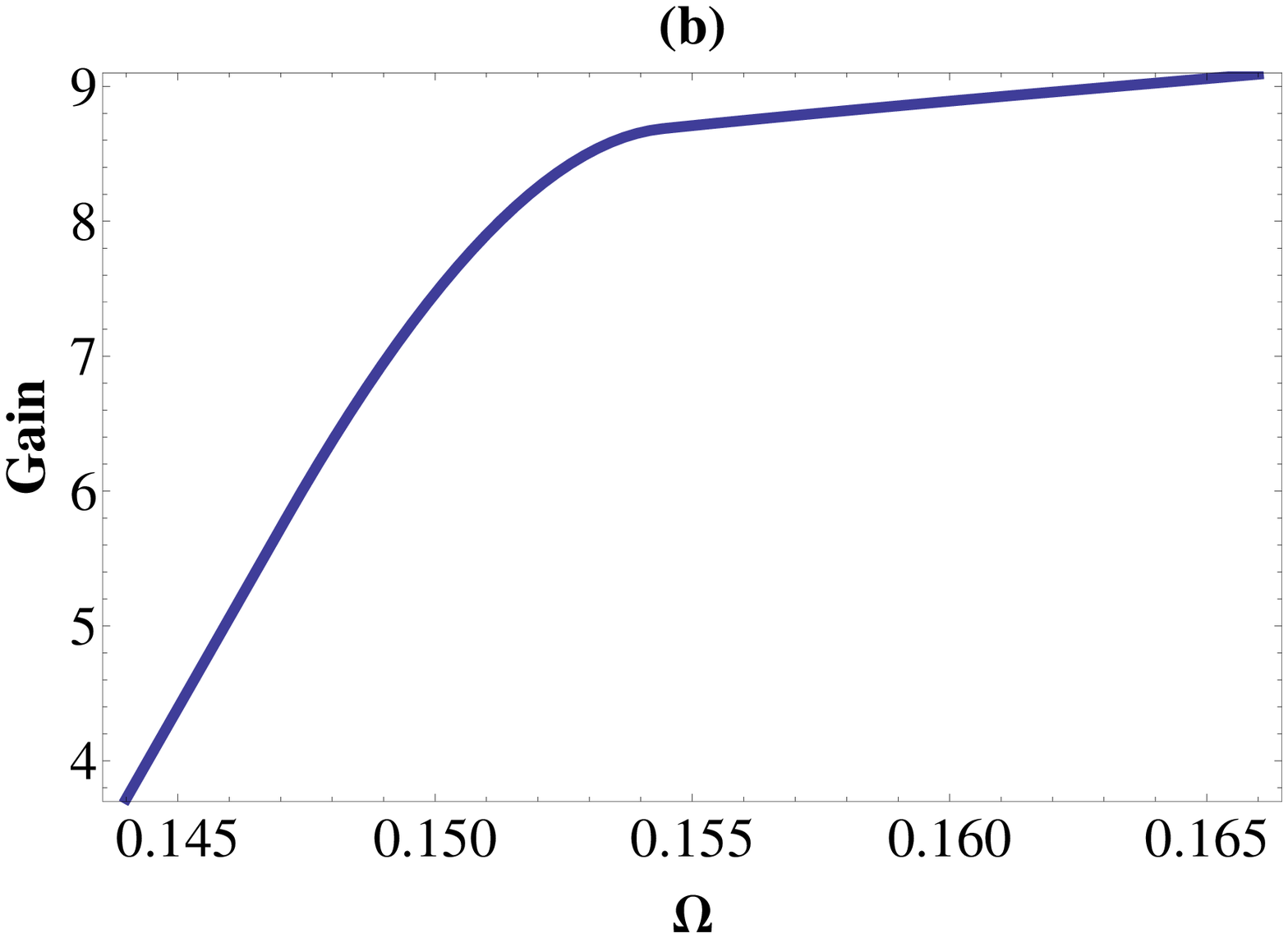} \\
	\end{tabular}
	\caption{The graph of (a)-Switching ratio v/s $P_{amp}$ for fix $\Omega$=1; (b)- Switching ratio v/s Modulation frequency for fix $P_{amp}=0.5$.  Parameters are used for graph are; J=0.5$\omega_{m}$, $\chi=0.3\omega_{m}$, $\kappa_{d}=1.8\omega_{m}$, N=0, g=$\omega_{m}$, $\theta=0.238\omega_{m}$, $\lambda$=0.02$\omega_{m}$, $\kappa_{a}=0.1\omega_{m}$,$\omega_{m}=1$, $\kappa_{b}=0.1\omega_{m}$, $\Delta_{a}$=1$\omega_{m}$, $\Delta_{b}$=1$\omega_{m}$, $\Delta_{d}$=0.}
\end{figure}

We find that by increasing $P_{amp}$, the gain decreases  while the bandwidth increases in Sinusoidal way for equal population in ground and excited state of quantum dot (i.e.,N=0). Initially the band width is lower for $P_{amp}$ as we increase its value the band width increases up to value 2.0$\omega_{m}$ and after that further increment of $P_{amp}$ band width slow down. This indicates that less transmission of output will occur for lower input signal i.e., "OFF" state for optical switch and as we increase the input signal the bandwidth is increasing, it leads to higher transmission of signal light i.e., "ON" state. Therefore, all-optical switching behavior is obtained. 

\begin{figure}[htb]
	\centering
	\begin{tabular}{@{}cccc@{}}
		\includegraphics[scale=0.4]{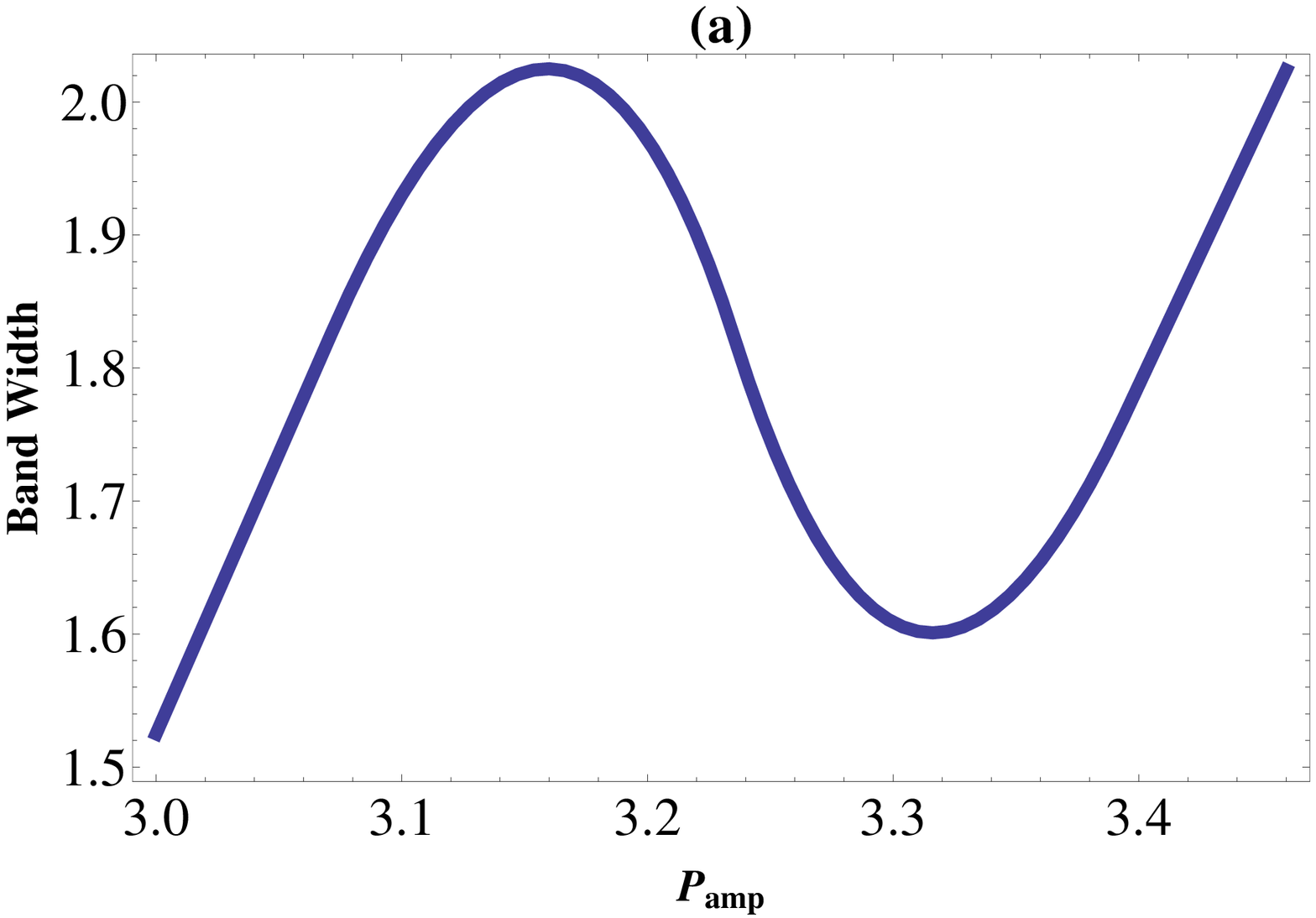}&
		\includegraphics[scale=0.4]{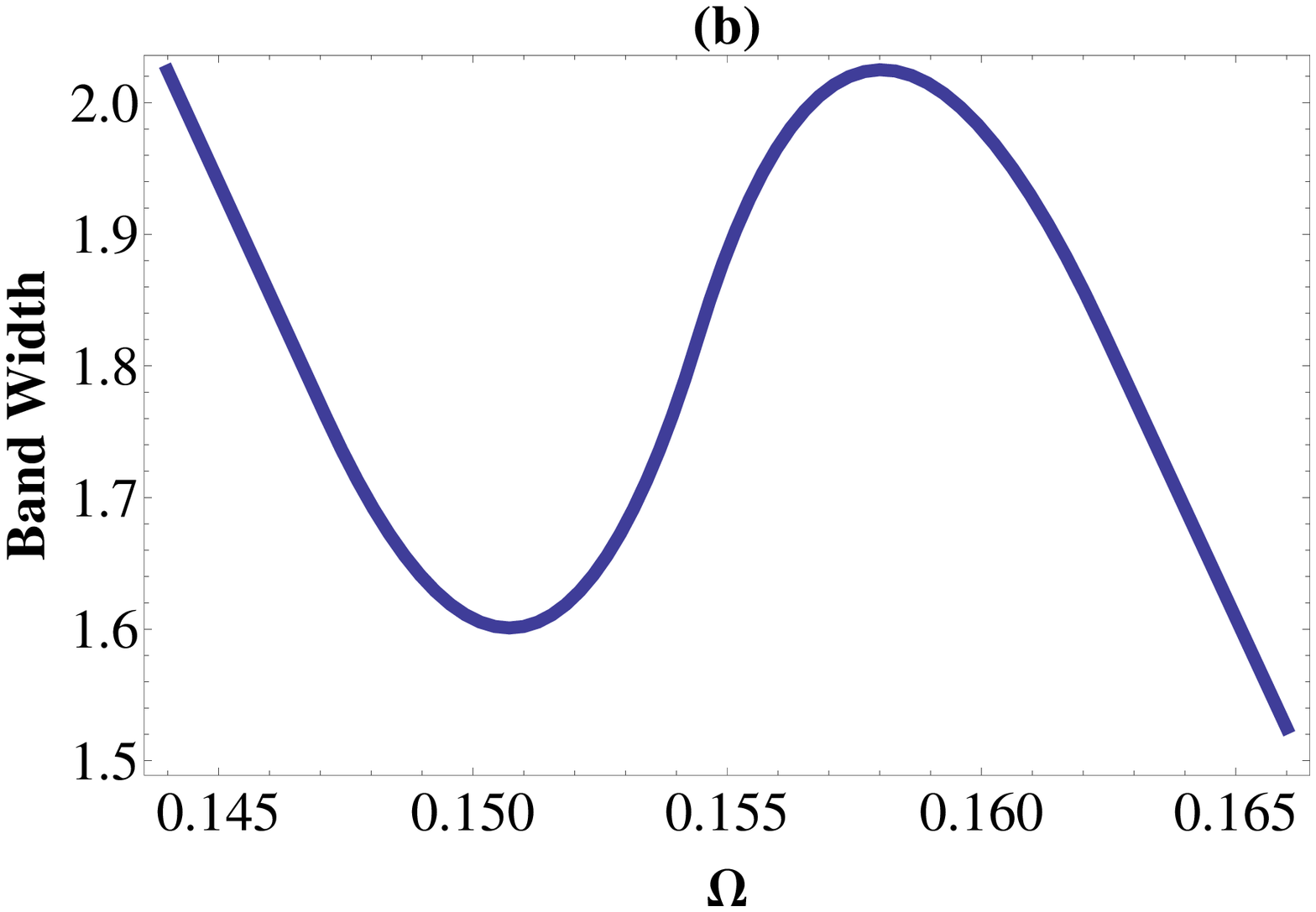} \\
	\end{tabular}
	\caption{The graph of (a)-Switching ratio v/s $P_{amp}$ for fix $\Omega$=1; (b)- Switching ratio v/s Modulation frequency for fix $P_{amp}=0.5$.  Parameters are used for graph are; J=0.5, $\chi=0.3\omega_{m}$, $\kappa_{d}=1.8\omega_{m}$, N=0,g=$\omega_{m}$, $\theta=0.238\omega_{m}$, $\lambda=0.02\omega_{m}$,$\omega_{m}=1$, $\kappa_{a}=0.1\omega_{m}$, $\kappa_{b}=0.1\omega_{m}$, $\Delta_{a}=1\omega_{m}$, $\Delta_{b}=1\omega_{m}$, $\Delta_{d}=0\omega_{m}$, $\eta$=0.1, C=0.36}
\end{figure}

\section{NORMAL MODE SPLITTING}

An interesting phenomenon occurs in the movable cavity mirror displacement spectrum known as NMS owing to the mixing of the system's various modes \citep{dob,bhatt,maha,sumei,tarun}. These spectrum characteristics of the probe field may prove further beneficial for achieving the optical switch in experiments and possible applications for coherent control of the light pulses\citep{wang}.

The NMS, is studied in this section to deduce the spectrum in presence of QD for small fluctuations in the movable DBR's position quadrature. Fourier transformation is used to convert the equation of motion(18-23) from the time domain to frequency domain for calculating the NMS. It is also possible to express these equations in condensed form $\dot{O}(t)$=MO(t) + f(t), where M being a 6$\times$6 time-independent matrix.

\[
M=
\begin{bmatrix}
0  &  \omega_{m} &  0  &  0  &  0  &  0  & \\
-\omega_{m} &  -\gamma_{m} &  0 &  0  &  a_{+}  &  -ia_{-}  & \\
0  &  0  &  -\kappa_{b}  &  \Delta_{b} &  0 &  J  & \\ 
0  &  0  &  -\Delta_{b}  &  -\kappa_{b} &  -J  &  0 & \\ 
ia_{-} &  0 &  0  &  J &  -\kappa_{a} &  \Delta & \\
a_{+} &  0 &  -J &  0 &  -\Delta &  -\kappa_{a} & \\
\end{bmatrix},
\]

 f(t)=[0,$\zeta$(t),$\sqrt{\kappa_{b}}u_{1in}(t)$,$\sqrt{\kappa_{b}}v_{1in}(t)$,$\sqrt{\kappa_{a}}u_{2in}(t)$,$\sqrt{\kappa_{a}}v_{2in}(t)]^T$ and
 
  $O(t) = [q(t),p(t),u_{1}(t),v_{1}(t),u_{2}(t),v_{2}(t)]^{T}$ are column vectors for noise sources and fluctuations respectively. Consequently displacement spectrum in Fourier space is given by,

\begin{equation}
S_{q}(\omega)= \frac{1}{4\pi}\int \left(<\delta q(\omega)\delta q(\omega^{'})+ <\delta q(\omega^{'})\delta q(\omega)>\right)e^{-i(\omega+\omega^{'})t}d\omega^{'}
\end{equation}

Now the system for Oscillating DBR displacement is obtained by applying correlation function(appendix A) in the Fourier space\citep{hua,shah,niel,pin,pat,bhatta,wipf,vital,vitali} as,

\begin{equation}
S_{q}(\omega)= \frac{1}{\left|Dd(\omega)\right|^2}\left(\left|K_{1}(\omega)\right|^2 + \left|K_{2}(\omega)\right|^2 + \left|K_{3}(\omega)\right|^2 + \left|K_{4}(\omega)\right|^2 + \left|K_{5}(\omega)\right|^2\right)
\end{equation}

Figure 6(a) illustrates movable DBR displacement spectrum $(S_{q}(\omega))$ versus frequency $(\omega/\omega_{m})$ (dimensionless). Three different values of coupling strength between two cavity modes produces displacement spectrum for J=0 (red solid line), J=1 (green dashed line) and J=1.5 (blue dash-dot line). Fig.6-(b) shows the  mechanical DBR's displacement spectrum $(S_{q}(\omega))$ against frequency $(\omega/\omega_{m})$ (dimensionless) for $\chi$= 0.2 (red solid line), $\chi$= 0.1 (green dashed line) as different optomechanical coupling values. We can see three peaks are arising in the NMS which are due to fluctuation of the two optical modes and coupling between the fluctuation of movable DBR's mechanical mode.

\begin{figure}[htb]
	\centering
	\begin{tabular}{@{}cccc@{}}
		\includegraphics[scale=0.4]{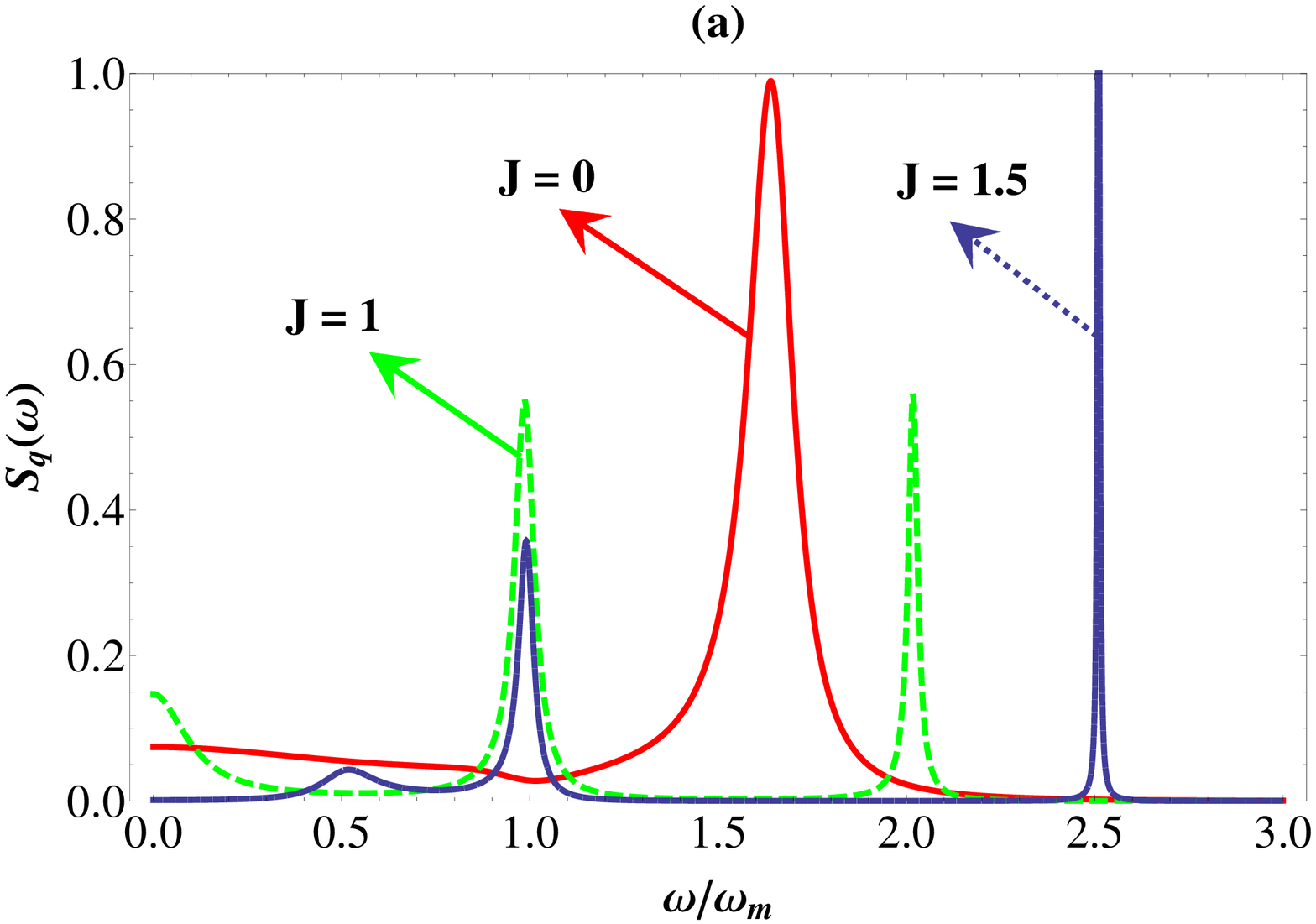}&
		\includegraphics[scale=0.4]{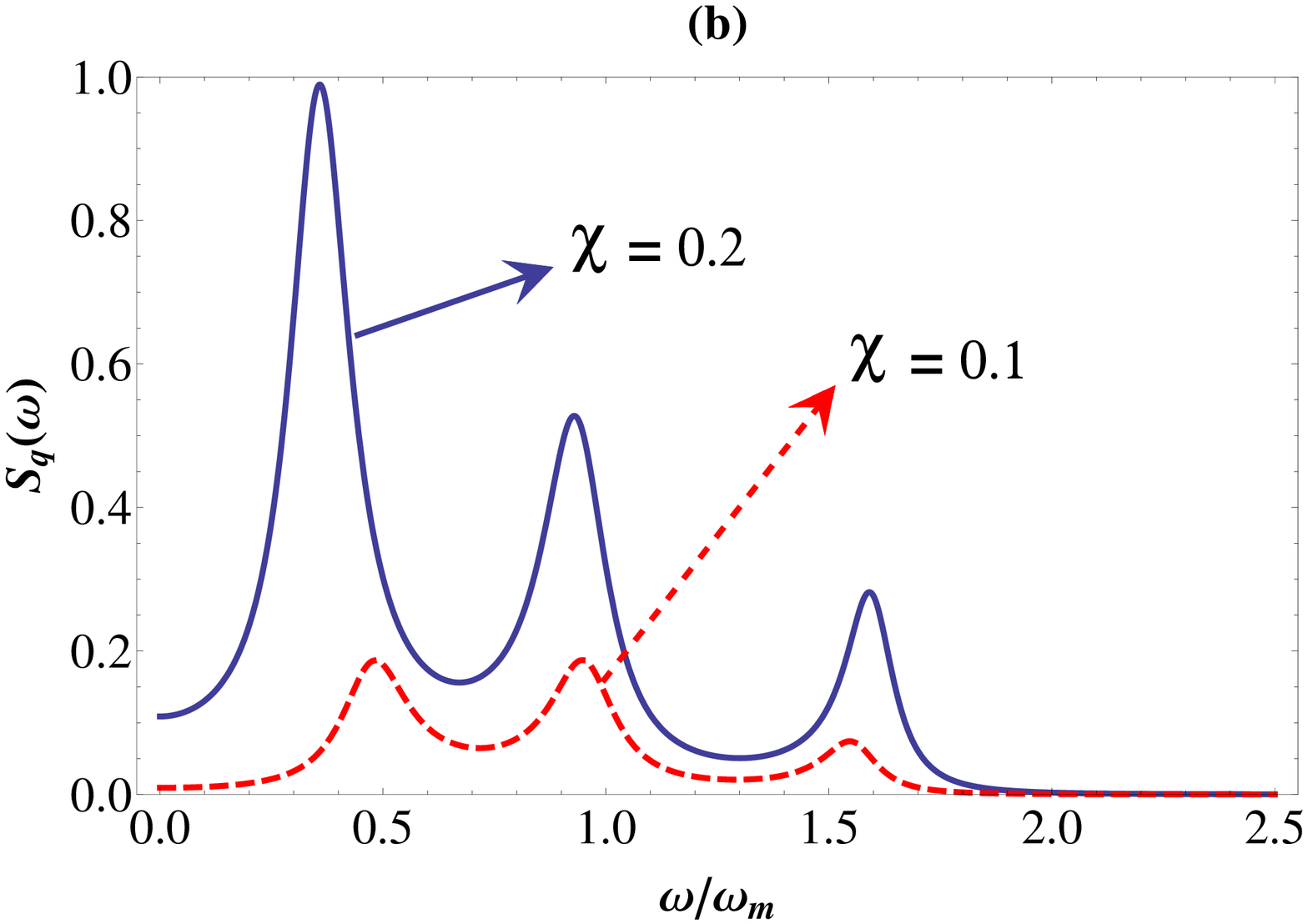} \\
	\end{tabular}
	\caption{The graph of Displacement Spectrum versus dimensionless frequency. In above figure, plot (a) displays the spectrum for varying coupling strength values between two cavity modes (J) for $\chi=0.2$ and plot (b) displays the spectrum for various optomechanical coupling constant values($\chi$) for J=0.5. Here, $\left(\frac{\hbar\omega_{m}}{\kappa_{b}T}\right)=10^{-6}$, $\kappa_{d}=1.8\omega_{m}$, N=0, g=1$\omega_{m}$, $\theta=0.238$, $\lambda=0.02\omega_{m}$, $\kappa_{a}=0.1$$\omega_{m}$, $\kappa_{b}=0.1$$\omega_{m}$, $\Delta_{a}$=1, $\Delta_{b}=1\omega_{m}$, $\Delta_{d}=-1\omega_{m}$, $\eta=0.1\omega_{m}$, C=0.10}
\end{figure}

The coherent energy exchange between the three modes can be observed. For NMS, each mode's coherence must be lower than energy exchange timescale between the three modes. Fig. 6-(b), shows that when $\chi=0.2$, then peaks have very high amplitude rather than for the value of $\chi$=0.1. This observation is because of the fact that there would be lesser energy swap between the optical and mechanical mode of cavity A if the optomechanical coupling is less. Thus the high amplitude peak shows the dominant behavior of energy swapping between the two optical modes.

\section{CONCLUSION}
 In this paper we have analyzed the bistable behavior of the optical switch and its performance. A hybrid optomechanical system is studied for its mechanical displacement spectrum, where it consists of two semiconductor microcavities coupled optically and containing a QD. The generated bistable behavior due to optomechanical  nonlinearity shows an optical switching behavior and it can be regulated by altering the laser power, rocking parameter, optomechanical coupling and  QD cavity coupling. A three peak NMS is seen due to the movable DBR displacement spectrum. The spectrum consisting of three peaks is a consequence from the exchange of energy among two optical and a mechanical mode. As revealed by the results, this system has the capacity to employed in sensitive optical switch and optical sensors. Two optically coupled cavity modes in which mechanical and optical modes are present show tunable entanglement by coherent exchange of energy. It shows that this type of hybrid optomechanical system in future can be used for transfer and storage of data signals and becomes a part of large quantum information processing unit.  
 
 \section{acknowledgements}
 	\textbf{P.K Jha} and \textbf{Vijay Bhatt} are thankful to \textbf{Department of Science and Technology DST(SERB), Project No. EMR/2017/001980, New Delhi} for the financial support. \textbf{S. A. Barbhuiya} acknowledges \textbf{BITS, Pilani} for the doctorate institute fellowship. \textbf{Aranya B. Bhattacherjee} is grateful to \textbf{BITS Pilani, Hyderabad campus} for the facilities to carry out this research.

\section{Appendix A}
 
The operators of input noise fulfill the following set of correlation functions \citep{vitali-2,man,gio,gio-2}

 \begin{equation}
\tag{A1}
 <a_{in}^{\dagger}(t) a_{in}(t')> =<a_{in}(t)a_{in}(t')> = 0,
\end{equation}

\begin{equation}
\tag{A2}
<b_{in}^{\dagger}(t)b_{in}(t')> = <b_{in}(t)b_{in}(t')> = 0,
\end{equation}

 \begin{equation}
\tag{A3}
<a_{in}(t)a_{in}^{\dagger}(t')> = \delta(t-t'),
\end{equation}

\begin{equation}
\tag{A4}
<b_{in}(t)b_{in}^{\dagger}(t')> = \delta(t-t').
\end{equation}

The following correlation function is satisfies by the mechanical mirror's Brownian force noise operator \citep{gio},

\begin{equation}
\tag{A5}
<\zeta(t)\zeta(t')> = \frac{\gamma_{m}}{2\pi\omega_{m}}\int \left[1+coth\left(\frac{\hbar\omega}{2\kappa_{b}T}\right)\right]\omega e^{-i\omega(t-t')}d\omega.
\end{equation}

where temperature of mechanical mirror-connected thermal bath is represented by T and $\kappa_{b}$ is Boltzmann Constant. Due to the connection between movable mirror and thermal bath, the mirror generates the Brownian noise due to random motion. In essence this sort of noise is Non-Markovian \citep{vital,man}.

The correlation function in Fourier space for various amplitude and phase noise quadrature along with the Brownian noise operator for displacement spectrum can be described as follows\citep{gio-2}:

\begin{equation}
\tag{A6}
<u_{2in}(\omega)u_{2in}(\Omega')> = 2\pi\delta(\omega + \Omega')
\end{equation}

\begin{equation}
\tag{A7}
<v_{2in}(\omega)u_{2in}(\Omega')> = -2i\pi\delta(\omega + \Omega')
\end{equation}

\begin{equation}
\tag{A8}
<u_{2in}(\omega)v_{2in}(\Omega')> = 2i\pi\delta(\omega + \Omega')
\end{equation}

\begin{equation}
\tag{A9}
<v_{2in}(\omega)v_{2in}(\Omega')> = 2\pi\delta(\omega + \Omega')
\end{equation}

\begin{equation}
\tag{A10}
<u_{2in}(\omega)u_{2in}(\Omega')> = 2\pi\delta(\omega + \Omega')
\end{equation}

\begin{equation}
\tag{A11}
<v_{1in}(\omega)u_{1in}(\Omega')> = -2i\pi\delta(\omega + \Omega')
\end{equation}

\begin{equation}
\tag{A12}
<u_{1in}(\omega)v_{1in}(\Omega')> = 2i\pi\delta(\omega + \Omega')
\end{equation}

\begin{equation}
\tag{A13}
<v_{1in}(\omega)v_{1in}(\Omega')> = 2\pi\delta(\omega + \Omega')
\end{equation}

\begin{equation}
\tag{A14}
<u_{1in}(\omega)u_{1in}(\Omega')> = 2\pi\delta(\omega + \Omega')
\end{equation}

\begin{equation}
\tag{A15}
<\zeta(\omega)\zeta(\Omega')> = 2\pi\frac{\gamma_{m}}{\omega_{m}}\omega\left[1+coth\left(\frac{\hbar\omega}{2\kappa_{b}T}\right)\right]\delta(\omega + \Omega').
\end{equation}

\section{Appendix B}
 
 The following are the unknown coefficients used in equation (26):
 
 \begin{eqnarray}
 Dd(\omega)&=&(-i\omega-\gamma_{m})\left(-iJ^{4}\omega + 2iJ^{2}\omega^{3} - i\omega^{5} + 2J^{2}\omega^{2}\kappa_{a} - 2\omega^{4}\kappa_{a} + i\omega^{3}\kappa_{a}^{2} + 2J^{2}\omega^{2}\kappa_{b} - 2\omega^{4}\kappa_{b}\hspace{3.5cm}\right.\\ 
 \nonumber
 && \left.-2iJ^{2}\omega\kappa_{a}\kappa_{b} + 4i\omega^{3}\kappa_{a}\kappa_{b} + 2\omega^{2}\kappa_{a}^{2}\kappa_{b} + i\omega^{3}\kappa_{b}^{2} + 2\omega^{2}\kappa_{a}\kappa_{b}^{2} - i\omega\kappa_{a}^{2}\kappa_{b}^{2} + i\omega^{3}\Delta^{2} + 2\omega^{2}\kappa_{b}\Delta^{2}\hspace{3cm} \right.\\ 
 \nonumber
 &&\left.-i\omega\kappa_{b}^{2}\Delta^2 + 2iJ^{2}\omega \Delta\Delta_{b} + i\omega^{3}\Delta_{b}^{2} + 2\omega^{2}\kappa_{a}\Delta_{b}^{2} - i\omega\kappa_{a}^{2}\Delta_{b}^{2} - i\omega\Delta^{2}\Delta_{b}^{2}\right)-\omega_{m}\left(-J^{4}\omega_{m} + 2J^{2}\omega^{2}\omega_{m}\hspace{1.89cm}\right. \\
 \nonumber &&\left.-\omega^{4}\omega_{m}-2iJ^{2}\omega\kappa_{a}\omega_{m}+2i\omega^{3}\kappa_{a}\omega_{m}+\omega^{2}\kappa_{a}^{2}\omega_{m}-2iJ^{2}\omega\kappa_{b}\omega_{m}+2i\omega^{3}\kappa_{b}\omega_{m}-2J^{2}\kappa_{a}\kappa_{b}\omega_{m}\hspace{1cm}\hspace{1.89cm}\right. \\ 
 \nonumber
 &&\left.+4\omega^{2}\kappa_{a}\kappa_{b}\omega_{m}-2i\omega\kappa_{a}^{2}\kappa_{b}\omega_{m}+\omega^{2}\kappa_{b}^{2}\omega_{m}-2i\omega\kappa_{a}\kappa_{b}^{2}\omega_{m}-\kappa_{a}^{2}\kappa_{b}^{2}\omega_{m} + \omega^{2}a_{-}^{2}\Delta-\omega^{2}a_{+}^{2}\Delta-2i\omega a_{-}^{2}\kappa_{b}\Delta \hspace{1.4cm}\right. \\
 \nonumber
 &&\left.+2i\omega a_{+}^{2}\kappa_{b}\Delta-a_{-}^{2}\kappa_{b}^{2}\Delta+a_{+}\kappa_{b}^{2}\Delta+\omega^{2}\omega_{m}\Delta^{2}-2i\omega\kappa_{b}\omega_{m}\Delta^{2}-\kappa_{b}^{2}\omega_{m}\Delta^{2}+J^{2}a_{-}^{2}\Delta_{b}^{2}-J^{2}a_{+}^{2}\Delta_{b} \hspace{2.2cm} \right. \\
 \nonumber
 &&\left.+2J^{2}\omega_{m}\Delta\Delta_{b}+\omega^{2}\omega_{m}\Delta_{b}^{2}-2i\omega\kappa_{a}\omega_{m}\Delta_{b}^{2}-\kappa_{a}^{2}\omega_{m}\Delta_{b}^{2}-a_{-}^{2}\Delta\Delta_{b}^{2}+a_{+}^{2}\Delta\Delta_{b}^{2}-\omega_{m}\Delta^{2}\Delta_{b}^{2}\right),
 \end{eqnarray}

\begin{eqnarray}
K_{1}(\omega)&=&\left(-J^{4}\omega_{m}+2J^{2}\omega^{2}\omega_{m}-\omega^{4}\omega_{m}-2iJ^{2}\omega\kappa_{a}\omega_{m}+2i\omega^{3}\kappa_{a}\omega_{m}+\omega^{2}\kappa_{a}^{2}\omega_{m}-2iJ^{2}\omega\kappa_{b}\omega_{m}+2i\omega^{3}\kappa_{b}\omega_{m}\hspace{1.7cm}\right.\\
\nonumber
&&\left.-2J^{2}\kappa_{a}\kappa_{b}\omega_{m}+4\omega^{2}\kappa_{a}\kappa_{b}\omega_{m}-2i\omega\kappa_{a}^{2}\kappa_{b}\omega_{m}
+\omega^{2}\kappa_{b}^{2}\omega_{m}-2i\omega\kappa_{a}\kappa_{b}^{2}\omega_{m}-\kappa_{a}^{2}\kappa_{b}^{2}\omega_{m}+\omega^{2}\omega_{m}\Delta^{2}\right.\hspace{2.4cm} \\
\nonumber
&&\left.-2i\omega\kappa_{b}\omega_{m}\Delta^{2}-\kappa_{b}^{2}\omega_{m}\Delta^{2}+2J^{2}\omega_{m}\Delta\Delta_{b}+\omega^{2}\omega_{m}\Delta_{b}^{2}-2i\omega\kappa_{a}\omega_{m}\Delta_{b}^{2}-\kappa_{a}^{2}\omega_{m}\Delta_{b}^{2}-\omega_{m}\Delta^{2}\Delta_{b}^{2}\right)\gamma_{m}Coth\left(\frac{\hbar\omega}{2K_{B}T}\right), 
\end{eqnarray}

\begin{eqnarray}
K_{2}(\omega)&=&(-iJ^{3}a_{-}\omega_{m}+iJ\omega^{2}a_{-}\omega_{m}+J\omega a_{-}\kappa_{a}\omega_{m}+J\omega a_{-}\kappa_{b}\omega{m}-iJa_{-}\kappa_{a}\kappa_{b}\omega_{m}+iJ\omega a_{+}\omega_{m}\Delta+Ja_{+}\kappa_{b}\omega_{m}\Delta\hspace{1.4cm} \\
\nonumber
&&+iJ\omega a_{+}\omega_{m}\Delta_{b}+Ja_{+}\kappa_{a}\omega_{m}\Delta_{b}+iJa_{-}\omega_{m}\Delta\Delta_{b})\sqrt{\kappa_{b}},\hspace{7.9cm} 
\end{eqnarray}

\begin{eqnarray}
K_{3}(\omega)&=&(-J^{3}a_{+}\omega_{m}+J\omega^{2}a_{+}\omega_{m}-iJ\omega a_{+}\kappa_{a}\omega_{m}-iJ\omega a_{+}\kappa_{b}\omega{m}-Ja_{+}\kappa_{a}\kappa_{b}\omega_{m}+J\omega a_{-}\omega_{m}\Delta-iJa_{-}\kappa_{b}\omega_{m}\Delta\hspace{1.3cm} \\
\nonumber
&&+J\omega a_{-}\omega_{m}\Delta_{b}-iJa_{-}\kappa_{a}\omega_{m}\Delta_{b}+Ja_{+}\omega_{m}\Delta\Delta_{b})\sqrt{\kappa_{b}}
\end{eqnarray}

\begin{eqnarray}
K_{4}(\omega)&=&\left(-iJ^{2}\omega a_{+}\omega_{m}+i\omega^{3}a_{+}\omega_{m}+\omega^{2}a_{+}\kappa_{a}\omega_{m}-J^{2}a_{+}\kappa_{b}\omega_{m}+2\omega^{2}a_{+}\kappa_{b}\omega_{m}-2i\omega a_{+}\kappa_{a}\kappa_{b}\omega_{m}-i\omega a_{+}\kappa_{b}^{2}\omega_{m}\right. \hspace{1.7cm}\\
\nonumber
&&\left.-a_{+}\kappa_{a}\kappa_{b}^{2}\omega_{m}+i\omega^{2}a_{-}\omega_{m}\Delta+2\omega a_{-}\kappa_{b}\omega_{m}\Delta-ia_{-}\kappa_{b}^2\omega_{m}\Delta+iJ^{2}a_{-}\omega_{m}\Delta_{b}-i\omega a_{+}\omega_{m}\Delta_{b}^{2}-a_{+}\kappa_{a}\omega_{m}\Delta_{b}^{2}\hspace{1.9cm}\right. \\
\nonumber
&&\left.-ia_{-}\omega_{m}\Delta\Delta_{b}^{2}\right)\sqrt{\kappa_{a}} \hspace{13.1cm} 
\end{eqnarray}

\begin{eqnarray}
K_{5}(\omega)&=&\left[\omega_{m}\left(a_{+}\left(-\omega^{2}\Delta+2i\omega\kappa_{b}\Delta+\kappa_{b}^{2}\Delta-J^{2}\Delta_{b}+\Delta\Delta_{b}^{2}\right)+ia_{-}\left(-J(iJ\omega+J\kappa_{b}) \right) \right) \right] \\
\nonumber
&&+\left[ \left ( \left(-i\omega-\kappa_{a})(-\omega^{2}+2i\omega\kappa_{b}+\kappa_{b}^{2}+\Delta_{b}^{2})\right)\right)\right]\sqrt{\kappa_{a}}
\end{eqnarray}


\begin{thebibliography}{99} 
 	
 	
 	\bibitem{saw}	
 	A. Sawchuk and T. C. Strand, “Digital optical computing,” Proceedings of the IEEE \textbf{72}, 758–779 (1984).
 	%
 	\bibitem{rao}
 	S. Rao, X. Hu, J. Xu and L. Li, J. Phys. B:At. Mol. Opt. Phys. \textbf{50} 055504 (2017).
 	%
 	\bibitem{gibbs}	
 	H. Gibbs, Optical Bistability: Controlling Light with Light (Academic Press, 1985).
 	%
 	\bibitem{mill}
 	D. A. B. Miller, “Are optical transistors the logical next step?” Nature Photonics 4, 3–5 (2010).
 	%
 	\bibitem{cau}
 	H. J. Caulfield and S. Dolev, “Why future supercomputing requires optics,” Nature Photonics 4, 261–263 (2014).
 	%
 	\bibitem{noz}
 	K. Nozaki, A. Shinya, S. Matsuo, Y. Suzaki, T. Segawa, T. Sato, Y. Kawaguchi, R. Takahashi, and M. Notomi, “Ultralow-power all-optical ram based on nanocavities,” Nature Photonics 6, 248252 (2012).
 	%
 	\bibitem{alm}
 	V. R. Almeida and M. Lipson, “Optical bistability on a silicon chip,” Opt. Lett. 29, 2387–2389 (2004).
 	%
 	\bibitem{sod}
 	M. Sodagar, M. Miri, A. A. Eftekhar, and A. Adibi, “Optical bistability in a one-dimensional photonic crystal resonator using a reverse-biased pn-junction,” Opt. Express 23, 2676–2685 (2015).
 	%
 	\bibitem{maj}
 	A. Majumdar and A. Rundquist, “Cavity-enabled self-electro-optic bistability in silicon photonics,” Opt. Lett. 39,
 	3864–3867 (2014).
 	%
 	\bibitem{kwon}
 	Y.-D. Kwon, M. A. Armen, and H. Mabuchi, “Femtojoule-scale all-optical latching and modulation via cavity nonlinear optics,” Phys. Rev. Lett. 111, 203002 (2013).
 	%
 	\bibitem{man1}
 	Q. Xu, S. Manipatruni, B. Schmidt, J. Shakya, and M. Lipson, “12.5 gbit/s carrierinjection-based silicon microring silicon modulators,” Opt. Express 15, 430–436 (2007).
 	%
 	\bibitem{liu}
 	A. Liu, R. Jones, L. Liao, D. Samara-Rubio, D. Rubin, O. Cohen, R. Nicolaescu, and M. Paniccia, “A high-speed
 	silicon optical modulator based on a metal-oxide-semiconductor capacitor,” Nature 427, 615–618 (2004).
 	%
 	\bibitem{vla}
 	Y. Vlasov, W. M. J. Green, , and F. Xia, “High-throughput silicon nanophotonic deflection switch for on-chip
 	optical networks,” Nat. Photonics 2, 242–246 (2008)
 	%
 	\bibitem{bri}
 	J. L. O’Brien, “ Optical quantum computing,” Science 318, 1567–1570 (2007)
 	%
 	\bibitem{pol}
 	A. Politi, M. J. Cryan, J. G. Rarity, S. Yu, and J. L. O’Brien, “ Silica-on-silicon waveguide quantum circuits,”
 	Science 320, 646–649 (2008).
 	%
 	\bibitem{eng1}
 	D. Englund, A. Faraon, B. Zhang, Y. Yamamoto, and J. Vuckovic,“ Generation and transfer of single photons on
 	a photonic crystal chip,” Opt. Express 15, 5550–5558 (2007).
 	%
 	\bibitem{far1}
 	A. Faraon, I. Fushman, D. Englund, N. Stoltz, P. Petroff, and J. Vuckovic,“ Dipole induced transparency in
 	waveguide coupled photonic crystal cavities,” Opt. Express 16, 12154–12162 (2008).
 	%
 	\bibitem{rossi}
 	M. Rossi, N. Kralj, S. Zippilli, R. Natali, A. Borrielli, G. Pandraud, E. Serra, G. Di Giuseppe and D. Vitali, Phys. Rev. Lett. \textbf{120}, 073601
 	%
 	\bibitem{thomp}
 	R. J. Thompson, G. Rempe, and H. J. Kimble, Phys. Rev. Lett. 68, 1132 (1992); C. Weisbuch et al., Phys. Rev. Lett. 69, 3314 (1992); A. Wallra et al., Nature (London) 431, 162 (2004); J. P. Reithmaier et al., Nature (London) 432, 197 (2004); T. Yoshie et al., Nature (London) 432, 200(2004); K. Hennessy et al., Nature (London) 445, 896 (2007).
 	%
 	\bibitem{wang}
 	T. Wang, M.-H. Zheng, C.-H. Bai, D.-Y. Wang, A.-D. Zhu, H.-F. Wang and S. Zhang, Ann. Phys. (Berlin) 2018, 1800228.
 	%
 	\bibitem{mari}
 	A. Mari and J. Eisert, "Gently modulating optomechanical systems," Phys. Rev. Lett. \textbf{103}, 213603 (2009).
 	%
 	\bibitem{valca}
 	G.J.de Valcarcel and K. Staliunas, Phys. Rev. E 67, 026604 (2003).
 	%
 	\bibitem{zhang}
 	X. Zhang, J. Sheng, H. Wu, Optic Express, \textbf{26}, 6285,2018.
 	%
 	\bibitem{gudat}
 	J. Gudat, "Cavity Quantum Electrodynamics with quantum dots in microcavities". PhD Thesis, University of Leiden (2012).
 	%
 	\bibitem{choy}
 	H.K.H. Choy, Design and fabrication of distributed Bragg reflectors for vertical-cavity surface-emitting lasers. M.Sc. Thesis, Mc Master University, 1996.
 	%
 	\bibitem{xu}
 	Q. Xu, S. Sandhu, M. L. Povinelli, J. Shakya, S.Fan and M. Lipson, Phy. Rev. Lett. 96, 123901 (2006).
 	%
 	\bibitem{cho}
 	J. Cho, D. G. Angelakis and S. Bose, Phys. Rev. A 78, 022323 (2008).
 	%
 	\bibitem{sato}
 	Y. Sato, Y. Tanaka, J. Upham, Y. Takahashi, T. Asano and S. Noda, Nat. Photon. 6, 56 (2012).
 	%
 	\bibitem{zheng}
 	C. Zheng, X. Jiang, S. Hua, L. Chang, G. Li, H. Fan and M. Xiao, Opt. Express 20, 18319 (2012).
 	%
 	\bibitem{peng}
 	B. Peng, S. K. Özdemir, F. Lei, F. Monifi, M.Gianfreda, G. L. Long, S. Fan, F. Nori, C. M. Bender and L. Yang, Nat. Phys.10, 394 (2014).
 	%
 	\bibitem{mark}
 	I. L. Markov, “Limits on fundamental limits to computation,” Nature \textbf{512}(7513), 147–154 (2014).
 	%
 	\bibitem{gosh}
 	M. Gosh, J. Phys. B:At. Mol. opt. Phys. \textbf{50} 165502 (2017)
 	%
 	\bibitem{chai}
 	Z. Chai, X. Hu, F. Wang, X. Niu, J. Xie and Q. Gong, Adv. Optical Mater, 1600665,(2016).
 	%
 	\bibitem{papa}
 	G. I. Papadimitriou, C. Papazoglou, A. S. Pomportsis, Journal of lightwave technology, \textbf{21}, 2, 2003.
 	%
 	\bibitem{dob}
 	J.M. Dobrindt, I. Wilson-Rae,  T.J. Kippenberg, Phys. Rev. Lett. 2008, \textbf{101}, 263602.
 	%
 	\bibitem{bhatt}
 	A.B. Bhattacherjee, Phys. Rev. A 2009, \textbf{80}, 043607.
 	%
 	\bibitem{maha}
 	S. Mahajan, N. Aggarwal, A.B. Bhattacherjee, Mohan,M.J. Phys. B: At. Mol. Opt. Phys. 2013, \textbf{46}, 085301.
 	%
 	\bibitem{sumei}
 	S. Huang, J. Physics. B: At. Mol. Opt. Phys. \textbf{47} 055504 (2014).
 	%
 	\bibitem{tarun}
 	T. Kumar, A.B. Bhattacherjee and ManMohan, "Dynamics of a movable micromirror in a nonlinear optical cavity," Phys. Rev. A \textbf{81}, 013835 (2010).
 	%
 	\bibitem{hua}
 	S. Huang,G.S. Agarwal, Phys. Rev. A 2009, \textbf{80}, 033807.
 	%
 	\bibitem{shah}
 	S. Shahidani, M.H. Naderi, M. Soltanolkotabi, S. BarzanjeH, J. Opt. Soc. Am. B 2014, \textbf{31}, 1087.
 	%
 	\bibitem{niel}
 	M.A. Nielsen, I.L. Chuang, Quantum Computation and Quantum Information; Cambridge University Press: Cambridge, 2000.
 	%
 	\bibitem{pin}
 	M. Pinard, A. Dantan, D. Vitali, O. Arcizet, T.Briant, A. Heidmann, Europhys. Lett. 2005, \textbf{72}, 747.
 	%
 	\bibitem{pat}
 	M. Paternostro, D. Vitali, S. Gigan, M.S. Kim, C. Brukner, J. Eisert, M. Aspelmeyer, Phys. Rev. Lett.2007, \textbf{99}, 250401.
 	%
 	\bibitem{bhatta}
 	M. Bhattacharya, P. Meystre, Phys. Rev. Lett. 2007, \textbf{99}, 073601; M. Bhattacharya, H. Uys, P. Meystre: Phys. Rev. A \textbf{77}, 033819 (2008).
 	%
 	\bibitem{wipf}
 	C. Wipf, T. Corbitt, Y. Chen, N. Mavalvala, New J. Phys. 2008, \textbf{10}, 095017.
 	%
 	\bibitem{vital}
 	D. Vitali, S. Gigan, A. Ferreira, H.R. Bhm, P. Tombesi, A. Guerreiro, V. Vedral, A. Zeilinger, M. Aspelmeyer,
 	Phys. Rev. Lett. 2007, \textbf{98}, 030405.
 	%
 	\bibitem{vitali}
 	D. Vitali, S. Mancini, P. Tombesi, J. Phys. A: Math. Theor. 2007, \textbf{40}, 8055.
 	%
 	\bibitem{vitali-2}
 	D. Vitali, S. Mancini, L. Ribichini, P. Tombesi, Phys. Rev. A 2002, \textbf{65}, 063803.
 	%
 	\bibitem{man}
 	S. Mancini, D. Vitali, P. Tombesi, Phys. Rev. Lett. 1998, \textbf{80}, 688.
 	%
 	\bibitem{gio}
 	V. Giovannetti, P. Tombesi, D. Vitali, Phys. Rev. A 1999, \textbf{60}, 1549.
 	%
 	\bibitem{gio-2}
 	V. Giovannetti, D. Vitali, Phys. Rev. A 2001, \textbf{63}, 023812.
 	%
 	


\end{thebibliography}
\end{document}